\documentstyle[buckow]{article}

\def\ee{\end{equation}}
\def\be{\begin{equation}}

\def\N{{\tt N}}

\begin{document}

\def\titleline{Counting BPS states of the M-5brane}
\def\authors{Giulio Bonelli\1ad}
\def\addresses{
\1ad
Physique Theorique et Mathematique - Universite Libre de Bruxelles\\
Campus Plaine C.P. 231 - B 1050 Bruxelles, Belgium\\
and\\
Theoretische Natuurkunde - Vrije Universiteit Brussel\\
Pleinlaan, 2 - B 1050 Brussel, Belgium\\
{\tt bonelli@phys.uu.nl}
}

\def\abstracttext{
An on-shell model for the supersymmetric index counting multiplicities
of BPS states of the M5-brane theory is reviewed. In particular we explicitly
study the tensionless Little String states appearing at
intersections in a bound state of $N$ 5-branes
wrapped on a six-manifold with product topology $M_4\times T^2$.
}
\large
\makefront

\section{Introduction}

A proper definition of M-theory as a non perturbative 
framework for superstring theories is still 
an open problem. It is strongly shaded, among other problems,
by a widely incomplete knowledge of
the very structure of the world-volume theory of the M5-branes.

A conjectural formulation of the M5-brane theory has been given in
\cite{open} in terms of a six-dimensional
string theory, called Little String Theory (LST), which should describe the
world-volume theory of the M5-brane. This arises naturally once we understand
the M-theory 5-brane not only as the magnetic dual of the membrane but also 
as Dirichlet surface for membranes. This picture is natural
from the point of view of the equivalence of M-theory compactification on a circle
$S^1$ and type IIA string theory. In fact, under $S^1$ compactification, 
the M5-brane
generates both the NS-5branes which are magnetic duals to fundamental strings
and the D4-branes on which fundamental strings can end. 
At the same time the fundamental
strings are generated by membranes wrapping the circle.

Despite the simplicity and naturalness of the above picture, 
LST shows several peculiar features which make its actual implementation
via the usual tools of superstring theory, to say the least, tricky.

A proper analysis \cite{bb} of the boundary degrees of freedom of
the membranes ending on the 5-brane gives as a result that 
the low energy spectrum of LST is given by the
$(0,2)$ self-dual tensor multiplet. Its bosonic content is given by 
a 2-form with self-dual curvature which couples minimally to the
strings and 5 bosons in the vector representation of the local
$SO(5)$ group describing the transverse dispalcements of the 5-brane in the 11
dimensional target space of M-theory (i.e. these 5 bosons are a section of the
normal bundle of the 5-brane world-volume embedded in 11 dimensions).

A first consequence of this analysis is that LST should not be understood 
in the usual perturbative terms. It is, because of self-duality, a symmetric dyon
with equal electric and magnetic charge and, applying the Dirac quantization
rule $eg'+ge'=2\pi n$, we find $e^2=\pi n$. This means that
the elementary charge $e_0=\sqrt{\pi}$ is not an adjustable parameter and therefore,
assuming that the charge is a non constant function of the string coupling,
it is natural to expect that LST does not admit
a usual {\it perturbative} world-sheet formulation.               
Another unusual feature is the LST target space dimension which is 6 and
this means that LST does not fulfill the criticality bound as a string theory.
Moreover a further possibly problematic aspect is given by its unusual 
non-gravitational low energy spectrum.

Notwithstanding the open questions concerning a proper {\it off-shell}
formulation of LST, it is possible to study some of its {\it on-shell} properties
by indirect methods and to check the validity of the above model.

The specific issue we are going to treat here is a counting of BPS multiplets
of states in LST. These are encoded in the generalized supersymmetric Witten
index of the theory. We will propose an on-shell framework to calculate
these objects within a six dimensional perspective on a world-volume six manifold 
of the product form $T^2\times M_4$.
The particular form of the six-manifold allows the use of a duality map
between M-theory on $T^2$ and type IIB theory in the limit of vanishing 
$T^2$-volume to check the validity of our results. 
Under this duality map, M5-branes wrapped on the above product
manifold correspond to D3-branes wrapped on $M_4$.
This map therefore reverts our problem to a calculation of a supersymmetric
path integral in a corresponding suitable four-dimensional gauge theory
whose coupling is identifyed with the $T^2$ modular parameter.

The result of our calculation will be to explicitly recognize the presence of 
tensionless string states in correspondence with the intersection between
different 5-branes in a given bound state and to give a precise formula to
count their multiplicities.

This talk is mainly based on \cite{io,io2}.

\section{The M5-branes index on $T^2\times M_4$}

To properly embed our problem in M-theory,
we consider M-theory on $W=Y_6\times T^2\times R^3$, where
$Y_6$ is a Calabi-Yau threefold of general holonomy.
Let $M_4$ be a supersymmetric simply connected four-cycle in $Y_6$
which we take to be smooth. Under these assumptions
$M_4$ is automatically equipped with a Kaehler form
$\omega$ induced from $Y_6$ and is simply connected.
We consider then $N$ M5-branes wrapped around $C=T^2\times M_4$.

It can be shown that in this specific geometrical set-up \cite{io2}
the potential anomalies which tend to ruin gauge invariance
of the world-volume theory are absent and that
it is then meaningful to define a supersymmetric index for the above 5-branes
bound states by extending the approach in \cite{vafa'}.
As it is well known, the supersymmetric index is independent on 
smooth continue parameters and
as a consequence we have that this counting of supersymmetry preserving 
states have to be independent on the $T^2$ volume.

In \cite{io2} the calculation of the supersymmetric index was performed 
in the limiting cases of small and of large volumes of the $T^2$
and shown to agree
while in \cite{io} a six dimensional framework was proposed which 
reproduces the above calculations as a one loop string supersymmetric 
path-integral. We will review the last construction.

Let us start with the single M5-brane case.
The bosonic spectrum of the low energy world-volume theory of this 5-brane is given by
a 2-form $V$ with self-dual curvature and five real bosons taking values
in the normal bundle $N_C$ induced by the structure of the embedding
as $T_W|_C=T_C\oplus N_C$. Passing to the holomorphic part and to 
the determinants and using the Calabi-Yau nature of $Y_6$,
it follows that the five transverse bosons are respectively, three non-compact 
real scalars $\phi_i$
and one complex section $\Phi$ of $K_{M_4}=\Lambda^{-2} T^{(1,0)}_{M_4}$,
which is the canonical line bundle of $M_4$.
A (partially) twisted chiral $(0,2)$ supersymmetry 
of the type considered in \cite{baulieuwest}
completes the spectrum.
It is given by a doublet of complex 
anti-commuting fields 
which are $(2,0)-$forms in six dimensions 
and a doublet of complex anti-commuting fields
which are scalars in six dimensions. 

The calculation of the supersymmetric index for this part of the spectrum 
can be done with path-integral techniques. It is classically exact
because of boson/fermion exact cancellation and it results in
a zero-modes amplitude for the self-dual tensor field.
This can be analyzed following the results of \cite{mans}.
It consists of a $\theta$-function of the lattice of the 
self-dual harmonic three forms.
The $\theta$-function is not completely specified in \cite{mans}
because of the possible 
inequivalent choices of its characteristics 
{\tiny $\left[\matrix{\alpha\cr\beta}\right]$}.
It is
$$\theta\left[\matrix{\alpha\cr\beta}\right](Z^0|0)
=\sum_k e^{i\pi \left((k+\alpha)Z^0(k+\alpha)+2(k+\alpha)\beta\right)},
$$ where $Z^0$ is a period 
matrix of the relevant six-manifold cohomology that we specify shortly.
Let $\left\{E^{(6)},\tilde E^{(6)}\right\}$ be a symplectic basis 
of harmonic 3-forms on the six-manifold at hand such that, in matrix notation,
$$
\int E^{(6)}E^{(6)}=0,\quad
\int \tilde E^{(6)}E^{(6)}=1,\quad
\int \tilde E^{(6)}\tilde E^{(6)}=0.
$$
We can expand $\tilde E^{(6)}=X^0E^{(6)}+Y^0{}^*E^{(6)}$, where ${}^*$ is the Hodge 
operator. Then $Z^0$ is defined as $Z^0=X^0+iY^0$.
In general, under modular transformations (i.e. global diffeomeorphisms of the
5-brane world-volume) these $\theta$-functions transform among each others.
If we ask for a single candidate closed under modular transformations, we see
\cite{gust}
that the natural choice is given by
{\tiny $\left[\matrix{\alpha\cr\beta}\right]=\left[\matrix{0 \cr 0}\right]$}.

In our case the world volume is in the product form $T^2\times M_4$
and \cite{io} we have 
$\frac{1}{2}b_3\left(T^2\times M_4\right)=b_2\left(M_4\right)$
and $Z^0=-\tau^{(1)}Q+i\tau^{(2)}1$, where $Q$ is the intersection matrix 
on $M_4$, $1$ is the unity matrix and $\tau=\tau^{(1)}+i\tau^{(2)}$ is the 
modulus of the $T^2$.
The relevant $\theta$-function is then
\be
\Theta(Z^0)\equiv\theta\left[\matrix{0 \cr 0}\right](Z^0|0)
= \sum_k e^{i\pi k Z^0 k}=
\sum_{m\in\Lambda}q^{\frac{1}{4}(m,*m-m)}{\bar q}^{\frac{1}{4}(m,*m+m)}
\equiv\theta_\Lambda(q,\bar q)
\label{uno}\ee
where $q=e^{2i\pi\tau}$ and $\Lambda$ is the lattice of integer period
elements in $H^2(M_4,{\bf R})$.

To count the full BPS spectrum of the theory a second sector is still lacking.
In fact, the 5-brane theory is completed in the UV by the little string theory
which has BPS saturates strings which eventually have to be kept into account
in the calculation of the complete supersymmetric index.
Even if a full off-shell model for this six-dimensional string theory is not 
available at the moment, it is possible to follow an on-shell simple
calculation scheme for their contribution to the supersymmetric index. 

As the susy index is given by a trace on the string Hilbert space, 
it corresponds to a one-loop string path integral. 
Moreover, as it is usual in these index calculations, 
the semiclassical approximation is exact.
Now, since $M_4$ is simply connected, the only contributions 
to this path integral can arise from string world sheets wrapping the $T^2$
target itself. 
Therefore the configuration space of $n$ of these string world-sheets 
\footnote{We are assuming here that only one type of BPS strings has to be
counted. This is supported also by the analysis of the short representation 
of $(0,2)$ supersymmetry in six dimensions given in \cite{agmh}.}
will be given
by the symmetric product $(M_4)^n/S_n$ whose points parametrize the transverse 
positions.
Now, since the supersymmetric index calculated the Euler characteristics
of the configuration space, we claim that the full contribution from these string BPS 
configurations is given by 
\be
q^{-\chi_{M_4}/24}
\sum_n q^n \chi\left(M_4^n/S_n\right)=
q^{-\chi_{M_4}/24}
\prod_{n>0}\frac{(1-q^n)^{b_{odd}}}
{(1-q^n)^{b_{even}}}=\eta(q)^{-\chi_{M_4}}
\label{due}\ee
where we fixed a global multiplicative factor $q^{-\chi_{M_4}/24}$ because of 
modularity requirements and we used well known results from \cite{vw}.
In \cite{io} it is shown that this result can be obtained also via 
a natural generalization of 
the construction done in \cite{bps5} for toroidal and K3 compactifications 
to the generic simply connected Kahler case.

The complete supersymmetric index for a single five-brane on $T^2\times M_4$
is then given multiplying the factor (\ref{uno}) and the factor (\ref{due})
and reads
$$
{\cal I}_1^{T^2\times M_4}=\frac{\theta_\Lambda}{\eta^\chi}\, .
$$

We now pass to the analysis of the multi 5-brane case.
We will start assuming that the BPS multi five-brane bound states are 
classified by holomorphic branched coverings of the relevant world 
volume manifold and by a choice of spin structure on it.
Notice that this will result in recostructing exactly the 
relevant supersymmetric index as it can be calculated from the 
zero $T^2$-volume limit. In this limit \cite{io,io2} we find in fact 
the four dimensional gauge theory corresponding to a dual multi D3-brane 
bound state in type IIB and the calculation is under full control
In \cite{io} also an independent duality argument is given
which supports this picture when the six manifold admits a regular $S^1$ 
fibration.
Suppose therefore that we are dealing with a BPS bound state of $N$ 5-branes 
wrapped on a six manifold ${\cal C}$. 
Then we have to consider the space of
rank N holomorphic coverings of ${\cal C}$ in the M-theory target space. 
Generically this space of coverings will be disconnected and reducible
into components consisting of connected irreducible coverings
of rank less or equal to $N$. Each of these irreducible holomorphic
coverings corresponds to an irreducible bound state.
Then the general structure formula for the supersymmetric index
of an irreducible bound state of $N$ 5-branes is given by the lifting to the 
spectral cover of the index formula.
In formulas, if ${\cal C}_N$ is a generic holomorphic covering of ${\cal C}$, 
then ${\cal C}_N=\cup_j {\cal C}_j^{irr}$,
where $\sum j=N$ and ${\cal C}_j^{irr}$ is irreducible.
The relative contribution to the index
is ${\cal I}_N^{\cal C}=\prod_j {\cal I}_1^{{\cal C}_j^{irr}}$.
This formula has to be understood still schematically since we 
didn't specified the sum over the spin structures.
In the case ${\cal C}=T^2\times M_4$, a generic irreducible holomorphic covering
is of the form $\tilde T^2\times \Sigma_r$, where $\tilde T^2$ is a 
torus with modulus $\tilde\tau=(a\tau+b)/d$ which covers $a\cdot d=n_r$ times 
the original $T^2$ and $d>b\geq 0$ and $\Sigma_r$ is a rank $r$ holomorphic
branched covering of $M_4$. The above multiplicities are of course constrained
by $n_r\cdot r=j$ which is the rank of the six dimensional irreducible covering 
${\cal C}_j^{irr}$ above.
The full supersymmetric index is then given by summing all along the whole
set of possible covering structures.
Let us perform the sum with respect to the $T^2$ coverings first.
To do it by preserving modularity, we
multiply each factor by a rescaling term $d^{-w-\bar w}$, where
$(w,\bar w)$ are the modular weights of ${\cal I}|_{n_r=1}$, and then we sum
over all the triples $(a,b,d)$ fulfilling the above condition.
This operation coincides exactly with the definition of the Hecke 
modular operator ${\cal H}_{n_r}$ and our partial sum now reads
$
{\cal I}_{(n_r,r)}^{T^2\times M_4}={\cal H}_{n_r}
{\cal I}_{(1,r)}^{T^2\times M_4}
$.
By including the sum over the appropriate spin structures and explicitating the 
dependence over the given irreducible rank $r$ covering $\Sigma_r$,
we finally get
\be
{\cal I}_{n_r,r}= {\cal H}_{n_r}
\sum_{\varepsilon}\frac{\theta_{\Lambda^{\Sigma_r}+x}}{\eta^{\chi_{\Sigma_r}}}
\label{eg}\ee
where $\varepsilon$ is a label for 
the square-roots of the canonical line bundle (spin structures) on $M_4$ 
with respect to a given one
as ${\cal O}_\varepsilon\otimes K^{1/2}$ with ${\cal O}_\varepsilon^2=1$,
$x=[{\cal O}_\varepsilon^{\otimes a+1}]$ shifts correspondingly the
lattice of integer periods $\Lambda^{\Sigma_a}$ on $H^2(\Sigma_a,R)$. 

As it was already mentioned in the introduction, the multi 5-brane 
interacting model we have in mind consists in a picture where the 
5-branes, beside being magnetic duals of the membranes, are also
Dirichlet hypersurfaces for them, the boundaries of the membranes 
being strings in the 5-brane world-volume.
In this picture one expects that tensionless BPS string states appear
when two or more 5-branes intersect each other.
In our geometric picture this is the branching locus of the covering 
six manifold.
Indeed the supersymmetric index (\ref{eg}) reveals clearly 
the apparing of these extra BPS string states.
In our particular case,
the $T^2$ holomorphic self-covering is unbranched and the little strings
contribution is explicitly exposed in the Dedekind $\eta$-function.
In fact it enters in the form 
$\left(\frac{1}{\eta}\right)^{\chi_{\Sigma_r}}$, where $\Sigma_r$ is a rank
$r$ 
holomorphic covering of $M_4$.
The explicit dependence on the branching locus or the covering $B_r$ appears
once we apply the Hurwitz formula
$\chi_{\Sigma_r}=r\cdot\chi_{M_4}-\chi_{B_r}$
which holds for the covering structure.
We find therefore that 
\be
\eta^{-\chi_{\Sigma_r}}=
\left[\eta^{-\chi_{M_4}}\right]^r
\cdot
\eta^{\chi_{B_r}}
\label{tre}\ee
We read this formula for multiplicities of massless BPS string states
as a superposition of the independent BPS states pertaining to each brane copy
which is corrected by a further term corresponding to the interacting theory.

\section{Conclusions and Open Problems}

In this talk we have reviewed a six dimensional framework
for the evaluation of the supersymmetric index of M-theory
five-branes on $T^2\times M_4$
based on a {\it on shell} model for BPS states in LST.
The appearing of extra massless BPS states in coincidence
with 5-branes intersection is clearly encoded 
as a by-product in our formulas.

As we explained in the previous section, the geometric model 
encoding the structure of the BPS bound states of 5-branes 
we refer to is based on the idea that the relevant informations 
about them are encoded in the (geometric) moduli space
of branched holomrphic coverings of the six manifold 
on which the system lives.
Although we have different arguments to justify this model,
a proof internal to the six dimensional interacting theory 
is still lacking (together with the {\it off-shell} model itself)
and it sounds -- see how these structures appear in D-brane physics 
\cite{DDD,bbtt} -- 
that the direct solution of this specific point will 
be available only once the 5-brane (non-local) analogous geometrical 
structure to the D-branes gauge bundle structure will be given.

Moreover, the above formula for the supersymmetric index (lets consider
for simplicity a single 5-brane) should generalize to
$$\Theta(Z^0_{\cal C}) / \N(Z^0_{\cal C})
$$
for a 5-brane on a generic supersymmetric six cycle ${\cal C}$
where ${\N(Z^0_{\cal C})}^{-1}$ has to be a modular function
of the period matrix $Z^0_{\cal C}$ of the six manifold
which reduces to $\eta(\tau)^{-\chi_{M_4}}$ in the case
${\cal C}=T^2\times M_4$ and which calculates the 
BPS multiplicities of LS states in the general case.

Another interesting point would be to generalize the above 
treatment to amplitudes of BPS (surface) operators 
of the type considered in \cite{surf}.

The interested reader is referred to \cite{io3} for further developments 
in the subjects treated here.

{\bf Acknowledgments}
I would like to warmly thank M.~Bertolini for interesting discussions,
the Departments of Physics of Padua and Turin Universities where these
results have been presented and the organizers of the Corfu' RTN meeting.
Work supported by the European Commission RTN programme
HPRN-CT-2000-00131 as subcontractor of Leuven University.

\end{document}